\documentclass[aps,pra,reprint,longbibliography,superscriptaddress]{revtex4-1}
\usepackage[utf8x]{inputenc}

\usepackage{mathtools}
\usepackage{amssymb}
\usepackage{braket}
\usepackage{siunitx}

\usepackage{hyperref}

\usepackage{graphicx}
\usepackage{color}

\begin{document}

\title{Numeric estimation of resource requirements for a practical polarization-frame alignment scheme for QKD}
\author{Brendon~L. Higgins}
\email{brendon.higgins@uwaterloo.ca}
\affiliation{Institute for Quantum Computing and Department of Physics and Astronomy, University of Waterloo, Waterloo, Ontario N2L~3G1, Canada}
\author{Jean-Philippe Bourgoin}
\affiliation{Institute for Quantum Computing and Department of Physics and Astronomy, University of Waterloo, Waterloo, Ontario N2L~3G1, Canada}
\affiliation{Current Affiliation: Aegis Quantum, Waterloo, ON, Canada}
\author{Thomas Jennewein}
\affiliation{Institute for Quantum Computing and Department of Physics and Astronomy, University of Waterloo, Waterloo, Ontario N2L~3G1, Canada}

\begin{abstract}
Due to physical orientations and birefringence effects, practical quantum information protocols utilizing optical polarization need to handle misalignment between preparation and measurement reference frames. For any such capable system, an important question is how many resources---e.g., measured single photons---are needed to reliably achieve alignment precision sufficient for the desired quantum protocol. Here we study the performance of a polarization-frame alignment scheme used in prior laboratory and field quantum key distribution (QKD) experiments by performing Monte Carlo numerical simulations. The scheme utilizes, to the extent possible, the same single-photon-level signals and measurements as for the QKD protocol being supported. Even with detector noise and imperfect sources, our analysis shows that only a small fraction of resources from the overall signal---a few hundred photon detections, in total---are required for good performance, restoring the state to better than \SI{99}{\percent} of its original quality.
\end{abstract}

\maketitle

\section{Introduction}

Quantum communications technologies promise to be exciting new avenues for disseminating, processing, and controlling information. The most commercially ready of these technologies, quantum key distribution (QKD), distills a secure encryption key from the measurement results of quantum states sent from one party, Alice, to another, Bob, via a quantum channel~\cite{Gisin2002, Scarani2009}. In particular, BB84~\cite{Bennett1984} and related QKD protocols utilize the no-cloning theorem on qubit states to guarantee that an eavesdropper cannot ascertain any bits of the key without introducing detectable noise into the measurement statistics.

Optical platforms are an obvious choice for communicating quantum information, and one common information carrier is the electromagnetic-field polarization of photonic states. However, optical polarization denotes a direction in space, and for many quantum communications protocols, such as polarization-encoded BB84, the relative alignment of Alice's and Bob's polarization reference frames is crucial to the protocol's serviceability. Furthermore, the phase of each transmitted state must also be preserved, posing a problem beyond spatial frame alignment for birefringent media such as optical fibers (see also Ref.~\cite{Jeffrey2006}).

Since 2009, our group at the Institute for Quantum Computing, University of Waterloo, has been developing long-distance QKD technology with the goal of achieving Earth-orbiting platforms that service an ecosystem of quantum-secured communications. This work is presently culminating in the QEYSSat quantum satellite mission being spearheaded by the Canadian Space Agency, announced in 2017~\cite{CSA2020}. Prior, we and colleagues conducted a number of proof-of-principle experiments demonstrating polarization-encoded BB84 in closely related contexts, including over high-loss channels~\cite{Bourgoin2015}, and moving platforms both terrestrial~\cite{Bourgoin2015b} and airborne~\cite{Pugh2017}.

In support of this work we developed a practical polarization alignment scheme which employs single-photon-counting tomographic characterization and optimized compensation to correct arbitrary polarization rotations and birefringence of the transmission channel. Here we describe the core quantum-mechanical operation of the scheme in detail, and analyze the scheme's performance to quantify the single-photon resources required for high-fidelity correction of the transmission channel's effect. Using Monte Carlo simulations, we show that excellent correction can be achieved using only a small fraction of the received photon detections, even in a realistic environment possessing noise and imperfect visibility intrinsic to the source.

\section{Polarization alignment}\label{sec:scheme}

Several frame-alignment schemes for quantum systems have been investigated with the aim of achieving the highest possible fidelity under particular constraints by employing multi-photon collectively-entangled states and/or measurements---see, e.g., Refs.~\cite{Bagan2000, Peres2001, Bagan2001a, Bagan2001b, Bagan2004, Ballester2004}. Implementing these is difficult in practice, and such schemes tend to scale poorly with distance due to losses acting independently on photons within these collective systems. A more practical approach is to augment quantum communications protocols to be reference-frame independent (RFI), such that they utilize polarization subspaces in a way that is insensitive to specific effects---e.g., utilizing the invariance of the circular polarization basis under physical rotations around the beam path (though not birefringence-induced phase effects) for QKD~\cite{Laing2010, DAmbrosio2012, Wabnig2013, WenYe2015}. Another approach applies a compensation based on measurements of a correlated side-channel, such as polarimetry of a strong, classical signal multiplexed into the fiber at times or wavelengths near (but distinguishable from) the quantum signal~\cite{Xavier2011, Sasaki2011}.

When creating the polarization alignment scheme described here, our desire was to avoid additional complexities by using the transmitted states and measurements that must already be present for BB84 QKD, to the extent possible. This amounts to single-photon states (or ``weak'' coherent states with photon numbers ${\lesssim}1$, as a good approximation) with polarizations selected from one of four options evenly spaced around the equator of the Poincar\'e sphere. This approach naturally assures correspondence between the probing states used for characterization and the states being employed by the communication protocol, assuming the time-variance of the channel's effect is relatively slow.

We first used this alignment scheme in the context of high-loss QKD experiments using a weak coherent pulse (WCP) source emitting at \SI{532}{\nm} wavelength~\cite{Bourgoin2015}. There it was employed so that gradual state deviations caused by temperature fluctuations in the source and optical fibers could be eliminated at the press of a button.

Following experiments used this WCP source to perform QKD from a transmitter on the roof of a building to a receiver in the bed of a moving truck~\cite{Bourgoin2015b}. There, the source was located in a temperature-controlled laboratory, producing states that were guided by \SI{\approx85}{\m} of single-mode optical fiber through the core of the building to a motorized pointing platform. In contrast, the receiver side consisted of free-space polarization analysis optics, mounted on a truck that remained essentially level throughout. Thus, Alice's lab-to-roof fiber was the biggest contributor to state deviation, owing to optical fiber temperature fluctuation and the motion of the pointing platform. For practicality, focus was put on compensating the fiber channel, by characterizing light arriving at the transmitter telescope via a pick-off and immediate ``Bob'' measurement---the free-space channel and truck receiver orientation effects were omitted. (Notably, because only the optical power exiting the transmitter telescope is relevant for QKD security, with this approach any power lost to the pick-off may be compensated by increasing the WCP source power.) The polarization alignment scheme itself was automated to operate once per second throughout the experiment.

With upgraded hardware---including larger telescopes, integrated receiver optics, and a faster, higher-quality source at \SI{785}{\nm} wavelength---a similar approach to polarization alignment was used to support a demonstration of QKD transmitted from the ground to an aircraft in flight~\cite{Pugh2017}. Most recently, that source was employed in a long-distance demonstration of time-bin encoded QKD~\cite{Jin2019}, where the polarization alignment scheme was used to stabilize polarization states for conversion into time-bin states.

In the following description of the scheme, a key assumption is that there are no significant polarization-sensitive non-unitary effects within the quantum channel. This is generally true of both atmospheric and fiber transmission, and any significant polarization-sensitive non-unitary effect would be inherently detrimental to QKD performance, regardless of frame alignment. For photons that successfully traverse the channel, the most significant effects on polarization---owing to birefringence and physical orientation---can then be characterized as $\text{SU}(2)$ rotations of the polarization-encoded qubit state. Any number of these effects can together be described as a quantum channel which applies a single $\text{SU}(2)$ unitary $\hat U$ to any state sent through the channel. The alignment scheme is then conceptually separated into two main tasks: (a) characterize the action of the channel such that it has sufficient information about $\hat U$ to then (b) determine, and subsequently implement, a compensation operation $\hat V$ such that the combined action of $\hat U$ and $\hat V$ closely approximates identity.

\begin{figure}
 \centering\includegraphics[scale=0.9]{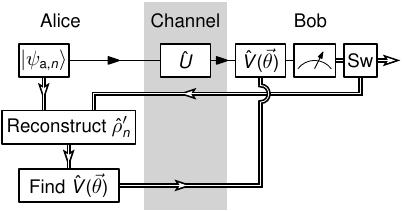}
 \caption{Schematic overview of the polarization alignment scheme as considered here. In this ``forward'' configuration, Alice prepares states $\ket{\psi_{\text{a}, n}}$ and transmits them to a receiver Bob, with the channel imparting an unknown unitary rotation $\hat U$. For characterization, Bob measures the states in a tomographically complete basis set---potentially making use of the controllable unitary $\hat{V}(\vec{\theta})$ to do so---and a logical data path switch (Sw) directs the results to Alice. (A switch could alternatively be placed prior to measurement, allowing unmeasured qubits to pass downstream, as in Refs.~\cite{Bourgoin2015b, Pugh2017, Jin2019}.) The detection coincidence counts are used to reconstruct the received states $\hat\rho_n'$ for each $n$. $\vec{\theta}$ are then optimized in a numerical model of compensation optics, $\hat{V}(\vec{\theta})$, and finally applied to the real apparatus for compensation of subsequent transmissions. A ``reversed'' configuration can be similarly constructed where $\hat{V}(\vec{\theta})$ is applied before the qubit traverses the quantum channel.}\label{fig:schematic_overview}
\end{figure}

\subsection{Characterization}

The effect of the quantum channel is characterized by transmitting a predefined set of states through the channel and analyzing measurement outcomes. Let $\ket{\psi_{\text{a},n}}$ be one such transmitted state, with $\ket{\psi_{\text{b},n}} = \hat U \ket{\psi_{\text{a},n}}$ being the corresponding state measured at the receiver.
Given a measurement eigenstate $\ket{\phi_m}$, the fidelity after applying the unknown unitary, $F(\phi_m | \psi_{\text{b},n}) = {\left|\braket{\phi_m|\psi_{\text{b},n}}\right|}^2$, quantifies the probability of obtaining that measurement outcome. We note that, in most practical situations, this is limited by an intrinsic signal fidelity, $F_\mathcal{S}$, owing to imperfections of source, channel, and measurement apparatuses leading to real or apparent depolarization. We can encompass these into an effective state $\hat{\rho}_n = (2F_\mathcal{S} - 1) \ket{\psi_{\text{b},n}} \bra{\psi_{\text{b},n}} + (1 - F_\mathcal{S}) \hat{I}$ being measured at the receiver after the unknown unitary of the channel is applied, with $\hat{I}$ being the identity operator.

To completely determine the action of $\hat U$, it is sufficient to accurately characterize the direction of one post-$\hat U$ state vector and the residual angular rotation about that vector. In practice we can satisfy this by transmitting ensembles of (at least) two different, non-orthogonal states (say, $\ket{\psi_{\text{a},1}}$ and $\ket{\psi_{\text{a},2}}$),
followed by single-qubit state tomography (i.e., single-photon polarimetry) of these states at the receiver.

We focus our attention on the $\ket H$, $\ket V$, $\ket D = (\ket H + \ket V)/\sqrt{2}$, and $\ket A = (\ket H - \ket V)/\sqrt{2}$ states prepared as part of the BB84 QKD protocol. If Bob's apparatus is sufficient to perform single-qubit state tomography, Alice may simply continue transmitting the same random sequence of states in $\ket H$, $\ket V$, $\ket D$, and $\ket A$, defining $\ket{\psi_{\text{a},n}}$ with $n \in \{1,2,3,4\}$, respectively. (Alternately, if Alice has an entangled photon source, as for the BBM92 protocol, she may measure her photon in order to project Bob's photon onto the $\ket H$, $\ket V$, $\ket D$, and $\ket A$ states---again just as she would do for the QKD protocol.) Temporal correlation of photon measurement events to the corresponding input states (i.e., particular values of $n$) can be performed using the same procedures Alice and Bob must already utilize for QKD. This allows photon measurement counts for each $n$ to be collected even though these states are sent in random order.

The four BB84 input states are more than necessary for complete characterization of $\hat U$---for example, the states $\ket H$ and $\ket D$ alone would be sufficient. The detection counts corresponding to the extra states orthogonal to each of these could be simply discarded, but in practice the measurement results are easily included in analysis and compensation (see below), and doing so improves the efficiency of the polarization alignment scheme with an unchanged QKD source.

Tomographic reconstruction of each state $\ket{\psi_{\text{b},n}} = \hat U \ket{\psi_{\text{a},n}}$ is then based on time-correlated photon detection statistics. Like the transmitted states, for practicality it makes most sense to utilize the measurement outcome eigenstates Bob already utilizes to implement BB84 ($\ket H$, $\ket V$, $\ket D$, and $\ket A$). Thus, for tomography, we use the tomographically complete set of measurements defined by the three Pauli matrices $\hat Z$ (projecting into $\ket H$/$\ket V$), $\hat X$ (projecting into $\ket D$/$\ket A$), and $\hat Y$ (projecting into $\ket R$/$\ket L$, where $\ket R = (\ket H + i\ket V)/\sqrt{2}$, and $\ket L = (\ket H - i\ket V)/\sqrt{2}$). In fact, this set of measurements is tomographically overcomplete, in the sense that they are more than necessary to extract full state information. Even so, measurements in the circular polarization basis, $\ket R$/$\ket L$, are necessary---states lying only on a great circle of the Poincar\'e sphere, such as BB84 states, are not tomographically complete. This places an additional requirement on Bob's apparatus beyond BB84 QKD. As we see below, the compensation mechanism can itself be utilized to achieve a change of basis necessary to implement these projections without further modifications to Bob's receiver.

The overcomplete basis set provides additional experimental robustness when compared to a complete basis set~\cite{Wootters1989} and, because most of these bases are also the BB84 measurement bases, can be practically implemented at the receiver. For each $n$, we tomographically reconstruct the density matrix $\hat\rho_n'$ from measured counts in these bases using maximum likelihood estimation~\cite{Hradil2000}. With these density matrices, an appropriate compensation can then be determined.

\subsection{Compensation}

We consider a compensation of the unitary $\hat U$ taking place at the receiver just prior to measurement. Any $\text{SU}(2)$ operation can be implemented in polarization optics by a \mbox{quarter-,} \mbox{half-,} quarter-wave plate arrangement, the operation being parametrized by the physical rotation of the three wave plates from their optic axes around the beam path. Determining the optimal compensation is thus a matter of optimizing the three wave plate orientation angles $\vec{\theta} = (\theta_1, \theta_2, \theta_3)$ such that each in the set of characterized states (each $\ket{\psi_{\text{b},n}}$) matches the corresponding transmitted state ($\ket{\psi_{\text{a},n}}$) with high fidelity.

For our implementation, we utilize the common Nelder--Mead simplex optimization algorithm. First, we construct a theoretical compensation unitary $\hat V(\vec{\theta}) = \hat Q(\theta_3) \hat H(\theta_2) \hat Q(\theta_1)$ encompassing the operation of the three wave plates given the parameters $\vec{\theta}$. The cost function of the algorithm, $C$, is then defined as the negative sum of fidelities between each state predicted after applying the compensation operation and the corresponding initial state---i.e.,
\begin{equation}
 C = -\sum_n \bra{\psi_{\text{a},n}} \hat V(\vec{\theta}) \hat{\rho}_n' \hat V^\dagger(\vec{\theta}) \ket{\psi_{\text{a},n}},
\end{equation}
here employing a more general form of fidelity to accommodate the density matrix $\hat{\rho}_n'$, which may not be a pure state.

If the reconstructed states $\hat\rho_n'$ accurately characterize the measured states $\ket{\psi_{\text{b},n}}$ (i.e., if $\hat\rho_n' \approx \hat U\ket{\psi_{\text{a},n}}\bra{\psi_{\text{a},n}}\hat U^\dagger$), then the cost function $C$ will be minimized when $\hat V(\vec{\theta}) \hat U = \hat I$. Minimizing $C$ by varying $\vec{\theta}$ thus optimizes the compensation operation. Applying the optimized theoretical wave plate orientations $\vec{\theta}$ to actual wave plates at the receiver implements the compensation and completes the alignment scheme, as pictured in Fig.~\ref{fig:schematic_overview}. (Although we do not directly reconstruct $\hat U$, in principle the inverse of $\hat V(\vec{\theta})$ will be a close approximation.)

Note that in the above formalism we have assumed, for simplicity, that while photon counting for characterization of $\hat U$ is taking place, the compensation wave plates are set such that they implement the identity $\hat I$ (e.g., by moving back to their optic axes). However, the effect of the wave plates not being at their optic axes during this phase, assuming their positions are known, can be straightforwardly incorporated. In addition, they could also be used for a secondary purpose if Bob's apparatus is only capable of photon counting in the $\hat Z$ and $\hat X$ bases. There, the compensation wave plates can be utilized to implement a change of basis prior to the state projection, and thereby achieve projections onto circular polarizations. (This could be done by, e.g., setting the second quarter-wave plate to an angle \SI{45}{\degree} from its optic axis, effectively transforming $\hat Z$ into $\hat Y$ for the remaining measurement time. This was done for Ref.~\cite{Bourgoin2015}.)

\subsection{Reversal using post-selection}

Typically, the compensation wave plates would be mounted in motorized rotation stages at the receiver, but for some situations it may be more suitable for these components to be placed at the transmitter. For example, such moving parts on an orbiting satellite platform introduce undesirable complexity and motion noise, making the scheme problematic for a satellite receiver platform~\cite{Bourgoin2013}. %
To address this, we exploit the time-symmetric nature of quantum mechanics to construct a ``reversed'' version of the above ``forward'' algorithm. Here, measurements of transmitted states are classified in a manner akin to post-selection, allowing us to establish an optimal pre-compensation operation that is applied to the photons immediately before leaving the transmitter.

Compared to the forward version of the scheme, in this reversed version the sets of input and measured states are swapped---for example, we define $\ket{\psi_{\text{a},n}} \in \{\ket H, \ket V, \ket D, \ket A, \ket R, \ket L\}$, and $\ket{\phi_m} \in \{\ket H, \ket V, \ket D, \ket A\}$. As before, measurement count statistics are collected for each combination of input and measured state. %
Let $d_{nm}$ be the counts for each input state index $n$ and measurement outcome index $m$. For each $m$, $d_{nm}$ covers a set of input states that forms a tomographically complete (in fact, overcomplete) basis set. By selecting the counts $d_{nm}$ for a fixed $m$ and performing tomography using those counts---i.e., over all the transmitted states, for each outcome---we reconstruct the effective input state conditional on post-selecting $\ket{\phi_m}$. In other words, we thus determine (up to the imprecision of the tomographic reconstruction) what states Alice would have sent Bob in order for them to become $\ket{\phi_m}$ upon application of the unknown unitary $\hat U$. The compensation is then optimized in the same manner as the forward scheme.

\section{Monte Carlo numerical simulations}\label{sec:sims}

To determine the performance characteristics of the alignment scheme, we conduct a series of numerical simulations incorporating a Haar-distributed random unitary $\hat U$, stochastic count generation, the characterization (tomography) and compensation (optimization) components of the scheme, virtual compensating wave plate operations, and resulting fidelity assessment. Using this, we perform a large number of Monte Carlo simulations of the scheme in both forward and reversed configurations.

For QKD, the primary goal is reducing the quantum bit error ratio (QBER), $E$, which quantifies the ratio of unexpected measurement outcomes to the total (within each basis relevant for QKD), and is used to determine the security of the channel~\cite{Scarani2009}. A lower QBER allows the bandwidth of key distribution to be increased, while a high QBER can cause the QKD protocol to be aborted with no secure key generated. The QBER is intimately related to the fidelities of the received states---specifically, $E = 1 - \sum_n F_n / 4$, where $F_n$ is the measured fidelity of the received state $\hat{\rho}_n$ against the expected $\ket{\phi_n}$. In the context of our compensation algorithm, it is straightforward to show that the QBER $E$ and cost function $C$ are linearly related, with minimal $C$ implying minimal $E$, so long as $\hat{\rho}'_n$ is an accurate estimate of $\hat{\rho}_n$.

\begin{figure}
 \centering\includegraphics{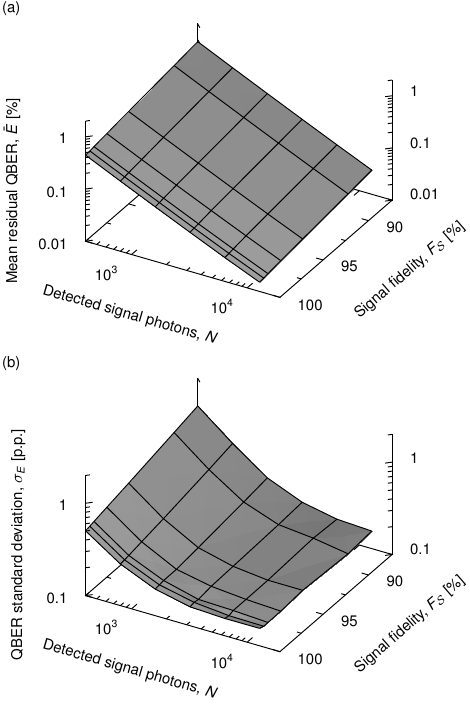}
 \caption{Performance of the forward polarization alignment scheme. (a) Mean residual QBER of nominal signal states after optimized compensation based on characterization analysis of $N$ detected signal photons with intrinsic signal fidelity $F_\mathcal{S}$ at the receiver. Only a few hundred photons are required to achieve low mean QBER. (b) Standard deviation of the mean residual QBER. Low photon counts and low intrinsic signal fidelities significantly increase the variation of performance between applications of the scheme.}\label{fig:forward_results}
\end{figure}

We quantify the polarization alignment scheme's performance from our simulation results using the mean predicted QBER, $\bar{E}$, of the nominal (ideal) signal states after the application of the channel unitary $\hat U$ followed by the compensation unitary $\hat V(\vec{\theta})$. With this definition, this ``residual'' QBER is zero for perfect compensation, regardless of the actual intrinsic signal fidelity $F_\mathcal{S}$.
We independently vary the number of detected photons $N$ and the intrinsic signal fidelity $F_\mathcal{S}$. Our results are calculated from $2^{24} \approx \SI[mode=text]{16.8}{\mega{} samples}$ of the Monte Carlo simulation (ensuring thorough convergence) for each of a total 468 configurations. From these results we obtain very good estimates of the expected mean residual QBERs and their standard deviations.

\subsection{Forward scheme}

The results for the forward scheme are illustrated in Fig.~\ref{fig:forward_results}(a). Given reception of a signal with perfect intrinsic fidelity, the mean residual QBER owing to the application of the polarization alignment scheme is less than \SI{0.5}{\percent} when at least $N = 400$ photons are measured (an average of 100 photons per input state, the lowest simulated), reducing to less than \SI{0.016}{\percent} for \num{12800} photons. (These values are consistent with common understanding that a few thousand copies are sufficient for producing good qubit state estimates via tomography~\cite{Rehacek2004}.) In other words, in this condition, as few as four hundred detections are sufficient to recover over \SI{99.5}{\percent} fidelity when an unknown unitary is acting on the channel.

The expected detection rate for a WCP source is $R [1 - (1 - Y_0) e^{-\eta \mu}]$, where $R$ is the source pulsing rate, $\mu$ is the coherent-state mean photon number, $\eta$ is the transmission of the channel, and $Y_0$ is the vacuum yield (per pulse) at the receiver. Neglecting the vacuum yield, for typical WCP source parameters $\mu = 0.5$ at $R = \SI{300}{\MHz}$, the expected detection rate over a channel with a significant \SI{40}{\dB} loss will be \SI{\approx 15}{\kHz}. Of this, 400 detections would be less than \SI{3}{\percent} of one second of data collection. This is less than some QKD implementations reveal publicly for the purpose of parameter estimation---such revealed outcomes could in fact be utilized also for polarization characterization. (Contrast this to correction techniques based on classical polarimetry, which utilize many orders of magnitude more photons.)

In the more realistic condition where the measured signal has imperfect intrinsic fidelity, the mean residual QBER increases---for example, for $F_\mathcal{S} = \SI{95}{\percent}$, 400 detections leads to \SI{0.59}{\percent} mean residual QBER, resulting from the inherent statistical uncertainty. A signal with at least $F_\mathcal{S} \approx \SI{87.5}{\percent}$ is necessary to maintain less than \SI{1}{\percent} mean residual QBER for 400 photons. Note that this intrinsic signal fidelity corresponds to an intrinsic signal QBER of at least \SI{12.5}{\percent}---too high to perform successful QKD, but evidently enough for good correction of $\hat U$. For comparison, good sources for QKD produce signals with fidelities in the vicinity of \SI{99}{\percent}. In the context of a satellite receiver, background and detector dark counts are expected to be the largest contributor to the imperfect intrinsic fidelity of the measured signal~\cite{Bourgoin2013}.

We perform a least-squared-error fit of the results of the simulation to a function of the form $\bar{E}(F_\mathcal{S}, N) = \alpha {(2F_\mathcal{S} - 1)}^\beta N^\gamma$. The optimized values, $\alpha \approx 2.93$, $\beta \approx -2.23$, and $\gamma \approx -1.07$, yield a coefficient of determination of \num{0.9999}, in excellent agreement with the data. In addition, the value of $\gamma$ corresponds quite well with the expected $1/N$ precision scaling of the tomographic reconstruction, suggesting that the tomography---necessary for characterizing the unitary---may be the limiting factor in the precision of the polarization alignment scheme.

\begin{figure}
 \centering\includegraphics{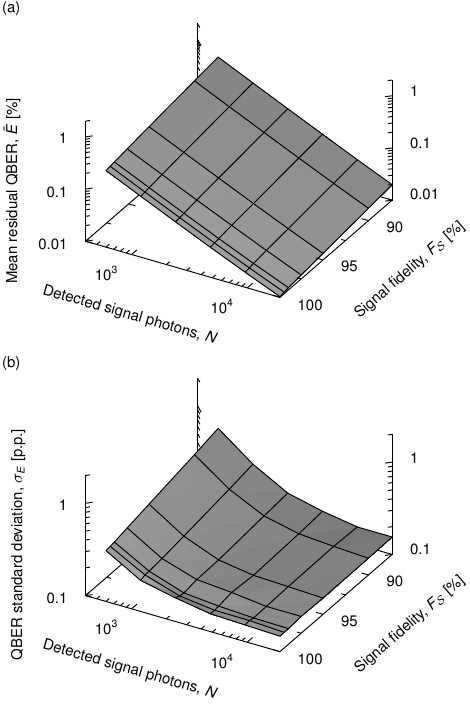}
 \caption{Performance of the reversed polarization alignment scheme. Subfigures (a) and (b) are as in Fig.~\ref{fig:forward_results}.}\label{fig:reverse_results}
\end{figure}

Figure~\ref{fig:forward_results}(b) shows the standard deviation of the residual QBER after compensation, illustrating the variability in the outcomes of each run. Given low numbers of photon detections, a drop in the intrinsic signal fidelity results in significant variations, exceeding one percentage point (p.p.)\ for $F_\mathcal{S} = \SI{87.5}{\percent}$ (and worse for lower $F_\mathcal{S}$ values not shown). With high intrinsic signal fidelity, however, the scheme behaves consistently, with standard deviations no more than about one-half of a percentage point.

\subsection{Reversed scheme}

The reversed scheme is also simulated, with results plotted in Fig.~\ref{fig:reverse_results}. As with the forward scheme, the general trend of better performance (lower mean residual QBER) with better intrinsic signal fidelity and higher numbers of detected photons is maintained. The overall performance, and variability, is very similar to the forward scheme. For example, with $N = 600$ measured photons (again, 100 photons per input state) and an intrinsic signal fidelity $F_\mathcal{S}$ of \SI{95}{\percent}, we find the reversed scheme achieves \SI{0.39}{\percent} mean residual QBER, comparable to the forward scheme. We again perform a least-squared-error fit to the function $\bar{E}(F_\mathcal{S}, N) = \alpha {(2F_\mathcal{S} - 1)}^\beta N^\gamma$, this time resulting in optimized values $\alpha \approx 3.25$, $\beta \approx -2.22$, and $\gamma \approx -1.08$, and yielding a coefficient of determination of \num{0.9998}.

\subsection{Impairment from background subtraction}

To try to improve the scheme under realistic use cases, we perform some additional simulations exploring the effect of simple background noise subtraction. For these simulations we add random (Poissonian) background counts to the simulated detected counts and then subtract the mean background (while guarding against unphysical negative counts) prior to the characterization step. The intent is to examine the effectiveness of subtracting from the measurements a known background level, perhaps determined in a calibration stage, to improve the signal-to-noise ratio.

\begin{figure*}
 \centering\includegraphics{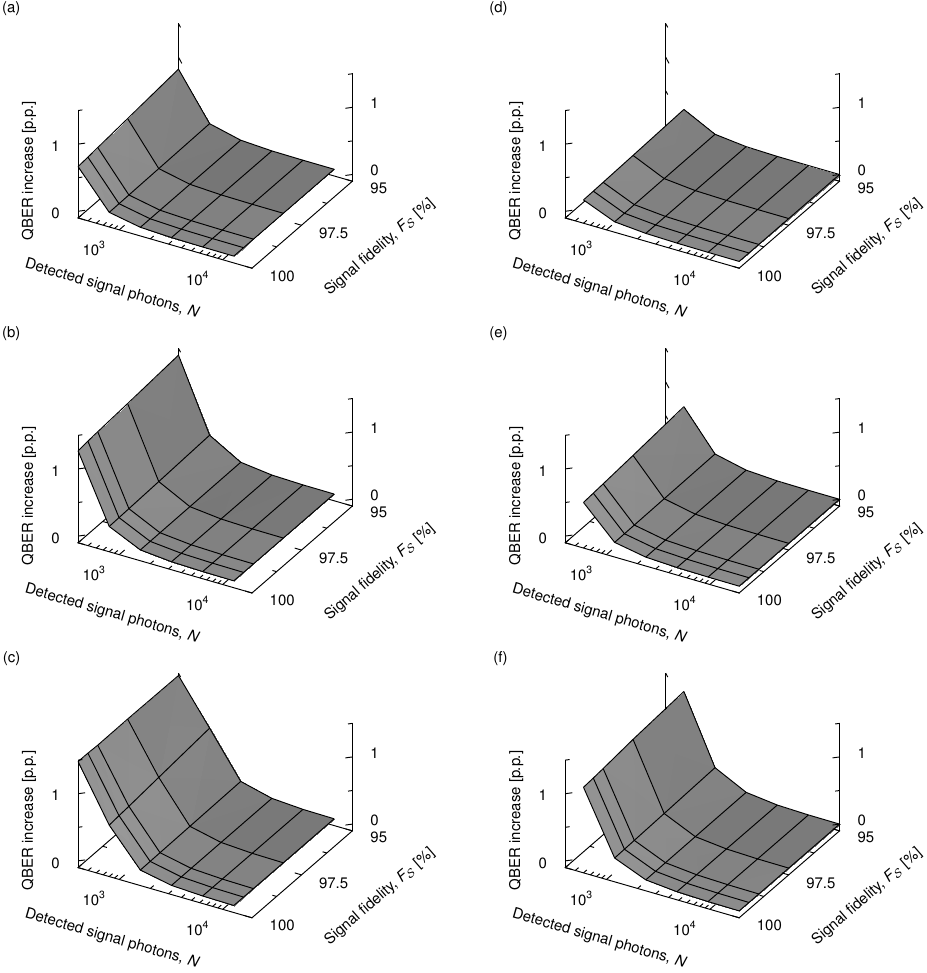}
 \caption{Increase of the mean residual QBER of the scheme when using background subtraction, as compared to without subtracting the background. The left column shows results for the forward case with (a) 100, (b) 200, and (c) 400 mean background counts in each of its six detectors. The right column shows results for the reversed case with (d) 100, (e) 200, and (f) 400 mean background counts in each of its four detectors. Nearly all cases show positive QBER increases, indicating worse outcomes when background is subtracted.}\label{fig:bg_diff}
\end{figure*}

For various photon detection counts $N$ and intrinsic signal fidelities $F_\mathcal{S}$, we compare the residual QBER of the alignment scheme operating with background subtraction, $\bar{E}_\text{BGS}$, against the alternative case where the background counts are added but the mean not subtracted, $\bar{E}_\text{BG}$. The results indicate that background subtraction is not better, and in many conditions clearly worse, than allowing the algorithm to operate using counts with background unaltered. This is illustrated by Fig.~\ref{fig:bg_diff}, which shows the increase of the residual QBER found when subtracting background (i.e., $\bar{E}_\text{BGS} - \bar{E}_\text{BG}$) in both forward and reversed cases.

\section{Discussion and conclusion}\label{sec:conclusion}

We have detailed a polarization frame alignment scheme tailored to BB84 QKD experiments, and theoretically characterized the photonic resources required for it to reliably obtain good alignment. The scheme utilizes only the single-photon-level states of the BB84 protocol itself and state generation or state measurement that is tomographically complete, which can be achieved by making intelligent use of the wave plates necessary to compensate the observed polarization rotation. Given less than a few percent of detection outcomes in a realistic QKD scenario, the residual QBER is below source intrinsic QBER, even over high-loss links---e.g., in any context where the signal fidelity is high enough to perform QKD, better than \SI{1}{\percent} residual QBER is possible with only 400 detections.

As a comparison, by maintaining optimal reference frame alignment, active polarization correction allows BB84 protocols to achieve greater secure key rates than RFI QKD in the general case~\cite{Wabnig2013}. Note also that RFI QKD assumes one basis remains well aligned, which is not in general true for $\text{SU(2)}$ unitary operations, and which the scheme presented here can naturally accommodate. Also interesting is that RFI QKD requires \emph{random} selection out of three bases for each qubit measurement, as this is a part of the security model, whereas a polarization alignment procedure which does not impact the security model (because it applies identical compensation to all states) can safely perform the \emph{same} measurement on multiple qubits in a row, for each basis. This supports flexibility for practical implementations.

It is clear that our polarization alignment scheme is not quantum-mechanically optimal---for ensembles of identical preparations, optimal approaches must use collective treatments~\cite{Massar1995}. It is, however, sufficiently efficient and relatively simple to implement as to make it practically useful. Interestingly, there is a trade-off between the number of photons repurposed from QKD key generation to perform the polarization alignment scheme, and the QBER achieved. An optimum must exist, as lower QBER increases the number of secure key bits that can be generated per received photon. However, many outcomes already revealed for parameter estimation could serve a dual role for polarization alignment characterization, mitigating the need for additional resources. This can also translate to other QKD schemes, such as measurement-device independent (MDI) QKD, where signals which would otherwise be discarded could instead be used for alignment.

More sophisticated techniques could be incorporated into the scheme to make it more theoretically efficient or practically compact. For example, direct estimation of state fidelities~\cite{Flammia2011} might provide a faster mechanism to assess whether full characterization and compensation is necessary. Or, where feasible, more optimal measurement approaches (e.g., Refs.~\cite{Rehacek2004, Ling2006}) or more compact polarimetery technologies (e.g., Ref.~\cite{Roy2016}) could potentially improve the practicality of the characterization apparatus. Other enhancements, such as estimators optimized for small-changes (e.g., Ref.~\cite{Kolenderski2008}) or online optimum-seeking control mechanisms (e.g., Refs.~\cite{Fisher2015, Ding2017}), with closed-loop control incorporating the compensation elements in the measurement, could be applied when deploying for continuous polarization control. In comparison to the continuous scheme described in Ref.~\cite{Ding2017}, which shows variable and sometimes very long settling times, such operation using the characterization and optimization approach presented here would compensate the channel effect quickly, accurately, and consistently.

\section{Acknowledgements}

We thank Nikolay Gigov for helpful discussions%
. We thank NSERC, Canadian Space Agency, CFI, CIFAR, Industry Canada, FedDev Ontario, and Ontario Research Fund for funding. B.L.H.\ acknowledges support from NSERC Banting Postdoctoral Fellowships (Canada).


\begin{thebibliography}{10}

\bibitem{Gisin2002}
N.~Gisin, G.~Ribordy, W.~Tittel, and H.~Zbinden.
\newblock Rev. Mod. Phys. 74, 145--195 (2002).

\bibitem{Scarani2009}
V.~Scarani, H.~Bechmann-Pasquinucci, N.~J. Cerf, M.~Du\v{s}ek,
  N.~L{\"u}tkenhaus, et~al.
\newblock Rev. Mod. Phys. 81, 1301--1350 (2009).

\bibitem{Bennett1984}
C.~H. Bennett and G.~Brassard.
\newblock In {Proceedings of the IEEE International Conference on Computers,
  Systems, and Signal Processing},  (Bangalore, India, 1984), pp. 175--179.

\bibitem{Jeffrey2006}
E.~R. Jeffrey, J.~B. Altepeter, M.~Colci, and P.~G. Kwiat.
\newblock Phys. Rev. Lett. 96, 150503 (2006).

\bibitem{CSA2020}
{Canadian Space Agency}.
\newblock {Quantum Encryption and Science Satellite (QEYSSat)}.
\newblock \url{https://asc-csa.gc.ca/eng/sciences/qeyssat.asp}.
\newblock Accessed 2020-04-21.

\bibitem{Bourgoin2015}
J.-P. Bourgoin, N.~Gigov, B.~L. Higgins, Z.~Yan, E.~Meyer-Scott, et~al.
\newblock Phys. Rev. A 92, 052339 (2015).

\bibitem{Bourgoin2015b}
J.-P. Bourgoin, B.~L. Higgins, N.~Gigov, C.~Holloway, C.~J. Pugh, et~al.
\newblock Opt. Express 23, 33437 (2015).

\bibitem{Pugh2017}
C.~J. Pugh, S.~Kaiser, J.-P. Bourgoin, J.~Jin, N.~Sultana, et~al.
\newblock Quantum Sci. Technol. 2, 024009 (2017).

\bibitem{Bagan2000}
E.~Bagan, M.~Baig, A.~Brey, R.~Mu{\~n}oz-Tapia, and R.~Tarrach.
\newblock Phys. Rev. Lett. 85, 5230--5233 (2000).

\bibitem{Peres2001}
A.~Peres and P.~F. Scudo.
\newblock Phys. Rev. Lett. 86, 4160--4162 (2001).

\bibitem{Bagan2001a}
E.~Bagan, M.~Baig, and R.~Mu{\~n}oz-Tapia.
\newblock Phys. Rev. A 64, 022305 (2001).

\bibitem{Bagan2001b}
E.~Bagan, M.~Baig, and R.~Mu{\~n}oz-Tapia.
\newblock Phys. Rev. Lett. 87, 257903 (2001).

\bibitem{Bagan2004}
E.~Bagan, M.~Baig, and R.~Mu{\~n}oz-Tapia.
\newblock Phys. Rev. A 70, 030301(R) (2004).

\bibitem{Ballester2004}
M.~A. Ballester.
\newblock Phys. Rev. A 69, 022303 (2004).

\bibitem{Laing2010}
A.~Laing, V.~Scarani, J.~G. Rarity, and J.~L. O'Brien.
\newblock Phys. Rev. A 82, 012304 (2010).

\bibitem{DAmbrosio2012}
V.~D'Ambrosio, E.~Nagali, S.~P. Walborn, L.~Aolita, S.~Slussarenko, et~al.
\newblock Nature Comm. 3, 961 (2012).

\bibitem{Wabnig2013}
J.~Wabnig, D.~Bitauld, H.~W. Li, A.~Laing, J.~L. O'Brien, et~al.
\newblock New Journal of Physics 15, 073001 (2013).

\bibitem{WenYe2015}
L.~Wen-Ye, W.~Hao, Y.~Zhen-Qiang, C.~Hua, L.~Hong-Wei, et~al.
\newblock Comm. Theor. Phys. 64, 295 (2015).

\bibitem{Xavier2011}
G.~B. Xavier, G.~{Vilela de Faria}, T.~{Ferreira da Silva}, G.~P. Tempor{\~a}o,
  and J.~P. von~der Weid.
\newblock Micro. Opt. Tech. Lett. 53, 2661--2665 (2011).

\bibitem{Sasaki2011}
M.~Sasaki, M.~Fujiwara, H.~Ishizuka, W.~Klaus, K.~Wakui, et~al.
\newblock Opt. Express 19, 10387--10409 (2011).

\bibitem{Jin2019}
J.~Jin, J.-P. Bourgoin, R.~Tannous, S.~Agne, C.~J. Pugh, et~al.
\newblock Opt. Express 27, 37214--37223 (2019).

\bibitem{Wootters1989}
W.~K. Wootters and B.~D. Fields.
\newblock Ann. Phys. 191, 363--381 (1989).

\bibitem{Hradil2000}
Z.~Hradil, J.~Summhammer, G.~Badurek, and H.~Rauch.
\newblock Phys. Rev. A 62, 014101 (2000).

\bibitem{Bourgoin2013}
J.-P. Bourgoin, E.~Meyer-Scott, B.~L. Higgins, B.~Helou, C.~Erven, et~al.
\newblock New J. Phys. 15, 023006 (2013).

\bibitem{Rehacek2004}
J.~\v{R}eh{\'a}\v{c}ek, B.-G. Englert, and D.~Kaszlikoski.
\newblock Phys. Rev. A 70, 052321 (2004).

\bibitem{Massar1995}
S.~Massar and S.~Popescu.
\newblock Phys. Rev. Lett. 74, 1259--1263 (1995).

\bibitem{Flammia2011}
S.~T. Flammia and Y.-K. Liu.
\newblock Phys. Rev. Lett. 106, 230501 (2011).

\bibitem{Ling2006}
A.~Ling, K.~P. Soh, A.~Lamas-Linares, and C.~Kurtsiefer.
\newblock J. Mod. Opt. 53, 1523--1528 (2006).

\bibitem{Roy2016}
S.~G. Roy, O.~M. Awartani, P.~Sen, B.~T. O'Connor, and M.~W. Kudenov.
\newblock Opt. Express 24, 14737--14747 (2016).

\bibitem{Kolenderski2008}
P.~Kolenderski and R.~Demkowicz-Dobrzanski.
\newblock Phys. Rev. A 78, 052333 (2008).

\bibitem{Fisher2015}
J.~Fisher, A.~Kodanev, and M.~Nazarathy.
\newblock J. Lightwave Technol. 33, 2146--2166 (2015).

\bibitem{Ding2017}
Y.-Y. Ding, W.~Chen, H.~Chen, C.~Wang, Y.-P. Li, et~al.
\newblock Opt. Lett. 42, 1023--1026 (2017).

\end{thebibliography}
\end{document}